\documentclass[showpacs,amsmath,amssymb,aps,pra]{revtex4}
\begin{document}

\title{A GENERAL FRAMEWORK FOR NONLINEAR QUANTUM DYNAMICS}
\author{S. Gheorghiu-Svirschevski\footnotemark[1]\footnotetext{e-mail: hnmg@soa.com}}
\address{1087 Beacon St., Suite 301, Newton, Massachusetts 02459}
\date{\today}


\begin{abstract}
We outline an approach that streamlines considerably the construction and analysis of well-behaved nonlinear quantum dynamics, with completely positive extensions to entangled systems. A few notes are added on the issue of quantum measurements under a nonlinear dynamics.    
\end{abstract}


\maketitle

\section{Introduction}
\label{Sec1}

In a previous paper \cite{qnlc-qd} we argued that the fundamental obstacle confronting nonlinear quantum theories is not an inherent incompatibility between nonlinear dynamics and the "no-signaling" condition, but the "probabilistic mixture" interpretation of mixed quantum states. If one upholds the "probabilistic mixture" point of view, the interpretation of measurement outcomes involves the "projection postulate", and particularly the "remote preparation" or "projection-at-distance" of entangled systems. Then according to an argument given in ref.\cite{Gisin}, quantum dynamics can only be linear. On the other hand, if all mixed states are regarded as "elementary mixtures" \cite{Bona-1}, in the sense applied to reduced local states corresponding to entangled pure states, the concept of "remote preparation" can no longer be supported, and dynamical nonlinearity becomes a theoretical possibility.

In support of this statement, we wish to sketch a straightforward framework for the formulation of relativistically well-behaved nonlinear dynamics, based on a technique applied in our previous work \cite{EQD} concerning a nonlinear entropic equation of motion \cite{Beretta-1,Beretta-2}. We consider only nonlinear extensions of the dynamical law, while preserving the usual "quantum statics" of states and observables, up to the interpretation of mixed quantum states as {\it elementary mixtures}. Furthermore, in view of our argument in ref.\cite{qnlc-qd}, we also demand that the limit of pure state dynamics be a linear, and eventually unitary, propagation. Our framework complements, and greatly simplifies, the more elaborate approach employed in a number of extensive works on nonlinear von Neumann equations \cite{Bona-1,Jordan,Czachor-1,Czachor-2,Czachor-3,Czachor-4,Czachor-5,Czachor-6,Czachor-7}. We provide as well an extended discussion of complete positivity in a nonlinear context \cite{Czachor-2}, and offer a few comments on a number of controversial points related to measurement theory.

\section{Physical prerequisites for a nonlinear dynamics}
\label{Sec2}
 
Let $\mathfrak H$ denote the Hilbert space of a closed [isolated] quantum system, let $L({\mathfrak H})$ be the linear space of linear operators on $\mathfrak H$, and $L_+ ({\mathfrak H})\subset L({\mathfrak H})$ be the convex set of positive definite, trace class density matrices $\hat\rho$. On general physical grounds, an application ${\bf g}_H : \hat\rho \to {\bf g}_H(\hat\rho)$ on $L_+ ({\mathfrak H})$, not necessarily linear, can represent a well-behaved dynamical map if: 

1) It conserves probability: $Tr[{\bf g}_H(\hat\rho)] = Tr[\hat\rho]=1$;

2) It is positive [${\bf g}_H(\hat\rho) \in L_+ ({\mathfrak H})$ for any $\hat\rho \in L_+ ({\mathfrak H})$] and remains positive in the presence of entanglement with a passive, noninteracting environment. This requirement seems naively equivalent to complete positivity, but the latter concept includes more then positivity constraints in a nonlinear context \cite{Czachor-2} [see also Sec.4 below].
 
3) It is local and separable, such that entangled mixed states that produce identical local initial conditions generate identical local evolutions, and uncorrelated states of noninteracting systems propagate into uncorrelated states. It will become apparent later in this Section that the separability condition is in fact insufficient for a correct characterization, and must be updated, e.g., to a condition of continuous and nonincreasing variation of the "degree of entanglement" of noninteracting systems.
 
\noindent To this we may add a very likely, and quite stringent, {\it pure state condition} \cite{qnlc-qd}:

4) All pure states of a closed [isolated] quantum system evolve in time according to a linear dynamics.

The problem of finding explicit nonlinear quantum dynamics that conform to these fundamental demands is generally approached in two steps: first, define a suitable class of positive, trace preserving dynamics for closed systems, and second, characterize its properties under external entanglement. Following the same philosophy, let us begin by considering the first task.

\section{Construction of positive and trace preserving dynamics}
\label{Sec3}  

We find that an easy way into this problem is provided by a square-root representation of density matrices as \cite{EQD}
\begin{equation}
\label{eq3}
\hat\rho = \hat\gamma \cdot \hat\gamma^{\dagger}\;,
\end{equation}
\noindent where the nonhermitian {\it state operator} [square root] $\hat\gamma \in L({\mathfrak H})$ has unit norm in the standard operator inner product $(\hat\alpha|\hat\beta) = Tr[\hat\alpha^{\dagger}\hat \beta]$. Any two $\hat\gamma$, $\hat\gamma'$ yielding the same density matrix $\hat\rho$ are related as $\hat\gamma' = \hat\gamma U$, with $U$ a unitary transformation, $UU^\dagger = U^\dagger U = I$. The task of finding {\it positive}, {\it trace preserving} maps ${\bf g}_H : \hat\rho \to {\bf g}_H(\hat\rho)$ on the convex set of density matrices $\hat\rho$ now becomes equivalent to finding {\it norm preserving} maps ${\bf u}_H : \hat\gamma \to {\bf u}_H(\hat\gamma)$ on $L({\mathfrak H})$, with the property that all $\hat\gamma$ corresponding to a unique $\hat\rho$ are mapped into ${\bf u}_H(\hat\gamma)$ corresponding to a unique ${\bf g}_H(\hat\rho)$. Note that maps ${\bf u}_H$ with this property automatically guarantee the positivity of the corresponding map ${\bf g}_H$ on density matrices. Note also that "overlap probabilities" $| (\hat\alpha|\hat\beta) |^2$ need not be conserved [in fact, do not have any physical meaning], and as a result there is no Wigner theorem to restrict the maps ${\bf u}$ to linear or antilinear applications. 

Let us begin by searching for maps that produce an evolution equation of the form 
\begin{equation}
\label{eq4}
i\hbar \dot{\hat\gamma} = G(\hat\gamma\cdot\hat\gamma^\dagger)\cdot \hat\gamma \;\;,
\end{equation}
\noindent where the generator $G$ is in general a nonhermitian operator that may have a nonlinear dependence on $\hat\gamma\cdot\hat\gamma^\dagger = \hat\rho$. Equations of this type are obviously invariant under transformations $\hat\gamma \to \hat\gamma U$, and generate well-defined equations of motion for the density matrix provided the norm $(\hat\gamma|\hat\gamma)  = Tr \hat\rho$ is conserved. If $G(\hat\rho)$ is represented as 

\[
G(\hat\rho) = T(\hat\rho) + i \Gamma (\hat\rho)\;,
\]

\noindent with $T = T^\dagger$ and $\Gamma = \Gamma^\dagger$, the density matrix equation of motion derived from Eq.(\ref{eq3}) reads
\begin{equation}
\label{eq5}
i\hbar \dot{\hat\rho} = \left[{ T(\hat\rho), \hat\rho }\right] + i \left\{ { \Gamma(\hat\rho), \hat\rho }\right\}\;\;,
\end{equation}
\noindent and the norm/trace conservation condition amounts to 
\begin{equation}
\label{eq6}
(\hat\gamma|{\bf \Gamma}|\hat\gamma)  = 0\;\;,
\end{equation}
\noindent where $\bf \Gamma$ denotes the superoperator defined by ${\bf \Gamma}\hat\gamma = \Gamma(\hat\gamma\cdot\hat\gamma^\dagger)\cdot \hat\gamma$. Should $\bf \Gamma$ be restricted to a linear application, Eq.(\ref{eq6}) would simply imply that ${\bf \Gamma} \equiv {\bf 0}$, and Eq.(\ref{eq5}) would reduce to the nonlinear von Neumann equations studied in refs. \cite{Jordan,Czachor-1,Czachor-2,Czachor-3,Czachor-4,Czachor-5,Czachor-6,Czachor-7}. Nevertheless, this is not the case when $\bf \Gamma$ is a nonlinear application. A typical counterexample is provided, e.g., by the superoperator 

\[
\bf \Gamma' \hat\gamma = \bf \Gamma \hat\gamma - \frac{(\hat\gamma|\bf\Gamma |\hat\gamma)}{(\hat\gamma|\hat\gamma)}\; \hat\gamma\;.
\]

\noindent Condition (\ref{eq6}) states that $\bf \Gamma$ must be in the class of {\it zero-mean superoperators}, which is a particular class of {\it fixed-mean superoperators}, satisfying $(\hat\gamma|{\bf \Gamma}|\hat\gamma)  = const.$ for any vector $\hat\gamma$ in their domain. 

Interestingly enough, the norm conservation condition (\ref{eq6}) also guarantees that the dynamics (\ref{eq5}) {\it always takes pure states to pure states}. To see this, it is sufficient to verify that demanding $(d/dt)(\hat\rho^2-\hat\rho) \equiv \dot{\hat\rho}\hat\rho + \hat\rho\dot{\hat\rho} - \dot{\hat\rho} = 0$ for $\hat\rho^2 = \hat\rho = |\psi\rangle \langle\psi|$ requires $\hat\rho \Gamma(\hat\rho) \hat\rho = 0$,  or equivalently, $\langle\psi| \;\Gamma(|\psi\rangle \langle\psi|)\; |\psi \rangle = 0$. But since a square-root $\hat\gamma$ for the pure state density matrix can only read $\hat\gamma = |\psi\rangle \langle\phi|$, with $|\phi \rangle$ an arbitrary normalized state vector, $\langle\phi|\phi \rangle =1$, it is seen that Eq.(\ref{eq6}) reduces to $\langle\psi|\;\Gamma(|\psi\rangle \langle\psi|)\; |\psi \rangle = 0$ as well. 

In general, the pure state evolution generated by Eq.(\ref{eq5}) is nonlinear. But there is a subclass of dynamics for which the right hand side of Eq.(\ref{eq5}) reduces to the familiar commutator form $[H, \hat\rho]$, with a unique linear and hermitian $H$, whenever $\hat\rho=\hat\rho^2$. In other words, there are nonlinear dynamics (\ref{eq5}) that comply with our {\it pure state condition}, and propagate {\it all} pure states in a linear and unitary fashion. For a nontrivial example consider 

\[
T(\hat\rho) = H\hat\rho^q + \hat\rho^q H\;,
\]

\noindent as in ref.\cite{Czachor-3}, and 

\[
\Gamma(\hat\rho) = \sigma \left[{ \hat\rho^r - \left({ Tr(\hat\rho^{r+1})/Tr(\hat\rho) }\right) }\right]\;,
\]

\noindent with $q$ and  $r$ positive scalars, and $\sigma$ a real, generally nonlinear functional of $\hat\rho$ . It is also possible to include supplementary conservation laws, e.g., energy conservation. For this case it suffices to redefine $\Gamma(\hat\rho) = \sigma \left[{ \hat\rho^r - \zeta H - \xi I }\right]$, where the scalars $\zeta$ and $\xi$ enter as Lagrange parameters to be determined from the norm and energy conservation conditions [see also Sec.VI of ref.\cite{EQD}]. Straightforward algebra will show that in both cases 

\[
[T(\hat\rho), \hat\rho] + i\{\Gamma(\hat\rho), \hat\rho \} = [H, \hat\rho] \;\;{\rm when} \;\;\hat\rho^2 = \hat\rho\;.
\]

The general solution for Eq.(\ref{eq4}) can be constructed in the form $\hat\gamma(t) = S_{\hat\rho}(t) \cdot \hat\gamma(0)$, where the $\hat\rho$-dependent propagator $S_{\hat\rho}(t)$ satisfies
\begin{equation}
\label{eq7}
i\hbar \dot{S}_{\hat\rho}(t) = G\left({ \hat\rho(t) }\right) \cdot S_{\hat\rho}(t)\;\;,
\end{equation}
\noindent and can be written symbolically as $S_{\hat\rho}(t) =  {\it  T} \exp\left[ { -(i/\hbar) \int\limits_0^t d\tau G\left({ \hat\rho(\tau) }\right) }\right]$. The associated solution for the density matrix reads then
\begin{equation}
\label{eq8}
\hat\rho(t) = S_{\hat\rho}(t) \hat\rho(0) S_{\hat\rho}^\dagger (t)\;\;,
\end{equation}
\noindent while the probability conservation condition becomes $Tr\left[{ S_{\hat\rho}(t) \hat\rho(0) S_{\hat\rho}^\dagger (t)}\right] = 1$ [similarly, energy conservation leads to $Tr\left[{ H S_{\hat\rho}(t) \hat\rho(0) S_{\hat\rho}^\dagger (t)}\right] = const.$].

\section{Extensions to entangled noninteracting systems: locality, separability and complete positivity}
\label{Sec4}

Let us consider now the behavior of a dynamics of type (\ref{eq8}) under external entanglement. We must begin with the observation that a quantitative refinement of properties (2) and (3) of physically well-behaved dynamics requires a proper extension of the nonlinear map ${\bf g}_H$ on $L_+ ({\mathfrak H})$ to a map conventionally denoted ${\bf g}_H \otimes {\bf I}_K$ on the set $L_+ ({\mathfrak H}\otimes {\mathfrak K})$ of density matrices over tensor product spaces ${\mathfrak H} \otimes {\mathfrak K}$, for arbitrary additional Hilbert spaces $\mathfrak K$. However, unlike the extensions of linear maps, which are uniquely defined by the very requirement of linearity, the extensions of nonlinear maps {\it are not uniquely defined}. 

For a more precise discussion, let $L({\mathfrak H}\otimes {\mathfrak K}) \equiv L({\mathfrak H})\otimes L({\mathfrak K})$ be the linear space of linear operators on ${\mathfrak H}\otimes {\mathfrak K}$, let ${\bf a}_H : {L({\mathfrak H})} \to {L({\mathfrak H})}$ be a linear application (superoperator), and let $\{\hat u_\alpha \}$ and $\{\hat v_\alpha \}$ be two basis sets in $L({\mathfrak H})$ and $L({\mathfrak K})$, respectively. To construct the linear extension $({\bf a}_H\otimes {\bf I}_K)$ of ${\bf a}_H$ onto $L({\mathfrak H})\otimes L({\mathfrak K})$ it is sufficient to define its action on the direct product basis $\{\hat u_\alpha \otimes \hat v_\beta \}$, as $({\bf a}_H\otimes {\bf I}_K) (\hat u_\alpha \otimes \hat v_\beta) = {\bf a}_H ( \hat u_\alpha) \otimes v_\beta$. Linearity then prescribes its action on any arbitrary element of its domain. One may be tempted for this reason to extend a similar definition onto nonlinear applications ${\bf g}_H : {L({\mathfrak H})} \to {L({\mathfrak H})}$ [or their restriction to $L_+ ({\mathfrak H})$], as done sometimes in the mathematical literature \cite{Ando-Choi,Anverson}. However, as pointed out in detail by Czachor and  Kuna \cite{Czachor-2}, the statement that $({\bf g}_H\otimes {\bf I}_K) (\hat u_\alpha \otimes \hat v_\beta) = {\bf g}_H ( \hat u_\alpha) \otimes v_\beta$ is far from sufficient to determine the extension uniquely. Moreover, the direct product mapping $\hat u_\alpha \otimes \hat v_\beta \to {\bf g}_H ( \hat u_\alpha) \otimes v_\beta$ will not survive a change of basis in either $L({\mathfrak H})$ or $L({\mathfrak K})$! 

To stress the severity of this statement, let us provide a rather striking example of {\it zero mean applications}. Let the linear superoperator ${\bf a}_H : {L({\mathfrak H})} \to {L({\mathfrak H})}$ be self-adjoint, and let the basis $\{\hat u_\alpha \}$ be its eigenbasis, for corresponding real eigenvalues $\lambda(\hat u_\alpha)$. Now define the {\it zero mean }application ${\bf a'}_H : {L({\mathfrak H})} \to {L({\mathfrak H})}$,

\begin{equation}
\label{eq9}
{\bf a'}_H(\hat w) = {\bf a}_H(\hat w) - \frac{Tr_H [\hat w^\dagger \cdot a(\hat w) ]}{ Tr_H [\hat w^\dagger \cdot \hat w]} \hat w \;\;.
\end{equation}  

\noindent This map has the remarkable property that it {\it vanishes identically on every element of the basis $\{\hat u_\alpha \}$}, since

\[
{\bf a'}_H(\hat u_\alpha) = {\bf a}_H(\hat u_\alpha) - \lambda(\hat u_\alpha) \hat u_\alpha \equiv 0\;\;.
\]

\noindent Nonetheless, ${\bf a'}_H$ is not a null application, and has nonvanishing values on a continuum of other elements of its domain, each of which can be expanded, of course, as a linear combination of the $\hat u_\alpha$-s. For instance, its action on any $\hat w = \mu \hat u_\alpha + \nu \hat u_\beta$, $\alpha \neq \beta$, amounts to

\[
{\bf a'}_H(\mu \hat u_\alpha + \nu \hat u_\beta) = \mu \; [\lambda(\hat u_\alpha) - \varepsilon] \; \hat u_\alpha + \nu \; [ \lambda(\hat u_\beta) - \varepsilon] \; \hat u_\beta \neq 0
\]

\noindent for 

\[
\varepsilon = \frac{ |\mu|^2 \;\lambda(\hat u_\alpha) \;Tr_H [\hat u_\alpha^\dagger \cdot \hat u_\alpha] + |\nu|^2 \; \lambda(\hat u_\beta) \;Tr_H [\hat u_\beta^\dagger \cdot \hat u_\beta] } {|\mu|^2 \;Tr_H [\hat u_\alpha^\dagger \cdot \hat u_\alpha] + |\nu|^2 \;Tr_H [\hat u_\beta^\dagger \cdot \hat u_\beta]} \;.
\] 

\noindent The same property carries over to an "extension" of the type $({\bf a'}_H\otimes {\bf I}_K) (\hat u_\alpha \otimes \hat v_\beta) = {\bf a'}_H ( \hat u_\alpha) \otimes \hat v_\beta$, but does not and cannot define the whole extension. This shows that the action of a direct product extension on product elements, $({\bf a'}_H \otimes {\bf I}_K)(\hat w \otimes \hat v_\beta) = {\bf a'}_H(\hat w) \otimes \hat v_\beta$ for all $\hat w \in {L({\mathfrak H})}$ and $\hat v_\beta \in {L({\mathfrak K})}$, must be prescribed in its  entirety in the definition of that extension. However, this prescription does not define the action of the desired extension on elements that are not of the product type, hence there exist an infinity of extensions with this same property.

From a physical point of view, the same prescription applied to a nonlinear quantum dynamical map ${\bf g}_H$ effectively enforces the {\it separability} of the extended map ${\bf g}_H \otimes {\bf I}_K$ on product states of noninteracting systems, i.e. $({\bf g}_H \otimes {\bf I}_K)(\hat w \otimes \hat v) = {\bf g}_H(\hat w) \otimes \hat v$ for all $\hat w \in {L({\mathfrak H})}$ and $\hat v \in {L({\mathfrak K})}$ {\it by definition}. But since this can no longer define the entire extension, it follows that one must necessarily use the other fundamental conditions on a well-behaved dynamics, positivity and locality, in order to select the proper extension, if one exists. It will be seen shortly that in fact there still remains a continuum of such extensions. One arrives in this way at the concept of a completely positive nonlinear extension put forth in refs.\cite{Czachor-1,Czachor-2}. We give it here a compact definition by employing the notions of {\it local equivalence class }and {\it positive local equivalence class }of a trace class product element of ${L({\mathfrak H}\otimes {\mathfrak K})}$:\\

{\it {\bf Definition 1} (local equivalence classes): Let $\hat\eta_H \in {L({\mathfrak H})}$ and $\hat\chi_K \in {L({\mathfrak K})}$ be trace class, such that $\hat\eta_H \otimes \hat\chi_K \in {L({\mathfrak H}\otimes {\mathfrak K})}$ is also trace class.

a) The set $\mathfrak{E}(\hat\eta_H \otimes \hat\chi_K)$ of all trace class elements $\hat w \in {L({\mathfrak H}\otimes {\mathfrak K})}$ with the property that $Tr_K [\hat w] = Tr_K [\hat \chi_K] \cdot \hat \eta_H$ and $Tr_H [\hat w] = Tr_H [\hat \eta_H]\cdot\hat \chi_K$ [such that $Tr_{H\otimes K} [\hat w] = Tr_{H\otimes K}[\hat\eta_H \otimes \hat\chi_K]$ ] defines the {\it local equivalence class }of the product $\hat\eta_H \otimes \hat\chi_K$. 

b) The {\it positive local equivalence class} of $\hat\eta_H \otimes \hat\chi_K$ is the set $\mathfrak{E}_+(\hat\eta_H \otimes \hat\chi_K)$ of all positive definite elements $\hat w$ in the {\it local equivalence class }$\mathfrak{E}(\hat\eta_H \otimes \hat\chi_K)$, $\mathfrak{E}_+(\hat\eta_H \otimes \hat\chi_K) =  {\mathfrak E}(\hat\eta_H \otimes \hat\chi_K)\cap {L_+ ({\mathfrak H}\otimes {\mathfrak K})}$.}\\

For the particular case when $\hat \rho_H \in L_+({\mathfrak H})$ and $\hat \rho_K \in L_+({\mathfrak K})$ are density matrices for two physical systems, $Tr_H[\hat \rho_H] = Tr_K[\hat \rho_K] = 1$, the {\it positive local equivalence class }of the uncorrelated state $\hat\rho_H \otimes \hat\rho_K$ contains the continuum of entangled mixed states [positive definite, unit trace density matrices] in ${L_+ ({\mathfrak H}\otimes {\mathfrak K})}$ that produce the local states $\hat \rho_H$ and $\hat \rho_K$. We are now in a position to give the following \\

{\it {\bf Definition 2} (completely positive extension of a nonlinear dynamical map): Let ${\bf g}_H :  L_({\mathfrak H}) \to L_({\mathfrak H})$ be a positive, trace preserving, generally nonlinear application. A completely positive extension of ${\bf g}_H$ onto a direct product space ${L ({\mathfrak H}\otimes {\mathfrak K})}$ is any application, denoted ${\bf g}_H \otimes {\bf I}_K$, with the property that for every $\hat \rho_H \in L_+({\mathfrak H})$, and $\hat \rho_K \in L_+({\mathfrak K})$, 

i) $({\bf g}_H \otimes {\bf I}_K) (\hat\rho_H \otimes \hat\rho_K) = {\bf g}_H(\hat\rho_H)\otimes\hat\rho_K$;

ii) $({\bf g}_H \otimes {\bf I}_K) \left( {\mathfrak E}_+(\hat\rho_H \otimes \hat\rho_K) \right) \subseteq {\mathfrak E}_+({\bf g}(\hat\rho_H) \otimes \hat\rho_K)$.}\\

From the definition of {\it local equivalence classes} it can be verified that every such extension is a positive, trace preserving, local and separable map on ${L({\mathfrak H}\otimes {\mathfrak K})}$. Also, for linear maps and linear extensions the above definition leads to the customary form employed in the standard linear theory. On the other hand, since ${\mathfrak E}_+(\hat\rho_H \otimes \hat\rho_K) $ is a continuum, it follows as well that there exists a continuum of completely positive extensions for every dynamics ${\bf g}_H$. Each of these extensions defines a distinct dynamics of entangled states. Moreover, it can be inferred that {\it there always exist completely positive nonlinear extensions of linear positive maps}. Hence a linear map that does not have linear completely positive extensions, such as the well-known transposition map, still has a continuum of nonlinear completely positive extensions. For the transposition map 

\[
{\bf t}_H(\hat\rho_H) = \hat \rho_H^T \;,
\]

\noindent a trivial example is given by 

\[
({\bf t}_H \otimes {\bf I}_K)_{nonlin}(\hat\rho_{H\otimes K}) = (Tr_K[\hat\rho_{H\otimes K}])^T \otimes Tr_H[\hat\rho_{H\otimes K}] \;.
\]

More generally, any map ${\bf g}_H$ can be trivially extended via 

\begin{equation}
\label{eq10} 
({\bf g}_H \otimes {\bf I}_K)_{nonlin}(\hat\rho_{H\otimes K}) = {\bf g}_H(Tr_K[\hat\rho_{H\otimes K}]) \otimes Tr_H[\hat\rho_{H\otimes K}]\;.
\end{equation}

\noindent But this trivial extension is obviously not physical, since it maps all entangled states into uncorrelated product states, and abruptly destroys all entanglement. This simple example shows that not all completely positive extensions of a [nonlinear] dynamics, as defined above, provide a physically acceptable dynamics of entangled systems. 

We are now prompted to recall the heuristic principle that local evolution should not increase the "degree of entanglement" of entangled noninteracting systems, and also that a continuous evolution in time should produce a "smooth" evolution of this "degree of entanglement". These "entanglement conditions" must be added explicitly to the characterization of physically well-behaved completely positive extensions. Obviously, a proper quantitative formulation of these requirements demands the introduction of a suitable measure of entanglement. 

However, if we limit our current purpose to a proof of existence of well-behaved extensions, this nontrivial problem can be circumscribed through a constructive approach that was first developed in refs.\cite{Jordan,Czachor-1,Czachor-2,Czachor-Doebner} for nonlinear von Neumann equations. Specifically, it suffices to show that a subset of dynamics described by Eq.(\ref{eq8}) admit so-called completely positive Polchinski extensions that comply with the above "entanglement principle". So consider a dynamics of type (\ref{eq8}) for a quantum system with Hilbert space $\mathfrak H$ entangled with a passive, noninteracting "environment" with Hilbert space $\mathfrak K$, such that the joint evolution reads

\begin{equation}
\label{eq11}
\hat\rho_{ H \otimes K }(t) = \left[{ S_{\hat\rho_{ H \otimes K }}^{ (H \otimes K) }(t) }\right] \hat\rho_{ H \otimes K }(0) \left[{ S_{ \hat\rho_{H \otimes K}}^{(H \otimes K)}(t) }\right]^\dagger \;\;.
\end{equation}

\noindent The sought extension is obtained by adapting Polchinski's conjecture \cite{Polchinski} and postulating the total propagator $S_{\hat\rho_{ H \otimes K }}^{ (H \otimes K) }(t) \in L({\mathfrak H}) \otimes L({\mathfrak K})$ as the direct product of local propagators for the local reduced states. If the "environment" has no internal dynamics, this means 
\begin{equation}
\label{eq12}
S_{\hat\rho_{ H \otimes K }}^{ (H \otimes K) }(t) {\mathop = \limits^{def}} S_{Tr_K \hat\rho_{H\otimes K}}^{(H)} (t) \otimes I_K\;,
\end{equation}
\noindent with $I_K$ the identity on $\mathfrak K$, and the extended dynamical map becomes 

\begin{equation}
\label{eq13}
\hat\rho_{ H \otimes K }(t) \equiv ({\bf g}_H \otimes {\bf I}_K )_t\left( \hat\rho_{ H \otimes K }(0)\right)=\left[{ S_{Tr_K \hat\rho_{H\otimes K}}^{(H)} (t) }\right] \hat\rho_{H\otimes K}(0) \left[{ S_{Tr_K \hat\rho_{H\otimes K}}^{(H)} (t) }\right]^\dagger .
\end{equation}

\noindent The differential form (\ref{eq4}) of this evolution is driven by a total generator

\begin{equation}
\label{eq14}
G^{({H\otimes K})} (\hat\rho_{H \otimes K} ) = G^{(H)} (Tr_K(\hat\rho_{H \otimes K}) ) \otimes I_K \;\;,
\end{equation}

\noindent where $G^{(H)}$ is again the local generator on $\mathfrak H$. It can be easily verified that for a noninteracting environment with a well-defined internal dynamics, the total dynamical map becomes, as usual, the composition [product] of commuting extensions of the local maps ${\bf g}_H$ and ${\bf g}_K$, 

\[
{\bf g}_{H\otimes K} \to {\bf g}_H \otimes {\bf g}_K \equiv ({\bf g}_H\otimes {\bf I}_K) \cdot ({\bf I}_H \otimes {\bf g}_K)\;\;,
\]

\noindent with a total generator given by the sum of local generators for the local states:

\[
G^{({H\otimes K})} = G^{(H)}(Tr_K(\hat\rho_{H\otimes K})) \otimes I_K + I_H \otimes G^{(K)}(Tr_H(\hat\rho_{H\otimes K}))
\]

\noindent The Polchinski-type extension defined above is known to be completely positive, in the sense of {\bf Definition 2}, for so-called nonlinear von Neumann dynamics \cite{Jordan,Czachor-1,Czachor-2}, with hermitian generators $G$ [$\Gamma=0$], and unitary propagators $S$. Let us provide here a theorem that clarifies the compatibility of such extensions with our more general framework, which allows for nonhermitian generators. As a preamble, it is convenient to introduce the concept of [non-]essential dissipative part for the generator of an evolution of type (\ref{eq5}).\\

{\it {\bf Definition 3} ([non-]essential dissipative part): The self-adjoint dissipative part $\Gamma^{(H)}(\hat\rho_H)$ of the generator $G^{(H)}(\hat\rho_H)$ is [non-]essential if for every $\hat\rho_H \in L_+ ({\mathfrak H})$ there does not exist any [there exists a] self-adjoint operator $W^{(H)}(\hat\rho_H)$ such that $\;i \{ \Gamma^{(H)}(\hat\rho_H), \;\hat\rho_H \} = [W^{(H)}(\hat\rho_H), \;\hat\rho_H]$.}\\

\noindent We will prove now \\

{\it {\bf Theorem  2}: A nonlinear dynamics of  type (\ref{eq5}) admits a completely positive Polchinski extension to noninteracting, entangled systems if and only if the nonhermitian part of its generator is non-essential.}  \\

{\it Proof: }We only need to prove the necessity of the generator form, but the following argument incidentally also covers sufficiency. Suppose that the generator of the evolution is nonhermitian. Complete positivity requires that the entangled dynamics must leave {\it both} the system and the environment locally unaffected [condition (ii) in {\bf Definition 2}]. Since definition (\ref{eq13}) for the extended dynamics already guarantees that the local system state remains unaffected, it follows that complete positivity demands that $\dot{\hat\rho}_K(t) \equiv Tr_H\dot{\hat\rho}_{H\otimes K}(t) = 0$ or, in view of Eq.(\ref{eq5}), that 
\begin{equation}
\label{eq15}
Tr_H\left[{\Gamma^{(H)}\left({ Tr_K \hat\rho_{H\otimes K} }\right) \cdot \hat\rho_{H\otimes K} }\right] = 0\;
\end{equation}
\noindent for any joint state $\hat\rho_{H\otimes K}(t)$. The meaning of Eq.(\ref{eq15}) becomes apparent with the observation that an entangled state $\hat\rho_{H\otimes K}$ cannot have nonzero matrix elements involving pure states $|\phi_H\rangle \in {\mathfrak H}$ outside the support of the corresponding local density matrix $\hat\rho_H = Tr_K(\hat\rho_{H\otimes K})$, i.e., such that $\hat\rho_H |\phi_H\rangle = 0$. The support of $\hat\rho_H$ is understood as the subspace subtended in $\mathfrak H$ by all eigenvectors of $\hat\rho_H$ with nonzero eigenvalues. This statement is trivially true for pure entangled states, where it is evident in the corresponding Schmidt decomposition. To prove its generalization for mixed entangled states, consider first the partial average $\langle \phi_H | \hat\rho_{H\otimes K} | \phi_H \rangle_{\mathfrak H} \in {\mathfrak K}$ for some arbitrary state vector $|\phi_H\rangle$ outside the support of $\hat\rho_H$, assuming any  exist. Since $\hat\rho_{H\otimes K}$ is positive definite, $\langle \phi_H | \hat\rho_{H\otimes K} | \phi_H \rangle_{\mathfrak H}$ should also be positive definite. But $Tr_K (\langle \phi_H | \hat\rho_{H\otimes K} | \phi_H \rangle_{\mathfrak H}) = \langle \phi_H | \hat\rho_H | \phi_H \rangle_{\mathfrak H}  = 0$, and so $\langle \phi_H | \hat\rho_{H\otimes K} | \phi_H \rangle_{\mathfrak H} = 0$. Let us now use the spectral decomposition $\hat\rho_{H\otimes K} = \sum\limits_\mu{|\mu_{H\otimes K}\rangle (\hat\rho_{H\otimes K})_{\mu\mu} \langle \mu_{H\otimes K}| }$  in the latter identity. Since $(\hat\rho_{H\otimes K})_{\mu\mu} \geq 0$, $\forall \mu$, it follows that $\langle \phi_H | \mu_{H\otimes K}\rangle_{\mathfrak H} = 0$ for all $\mu$ and any $|\phi_H\rangle$ outside the support of $\hat\rho_H$. This implies, in turn, that $\langle \phi'_H | \hat\rho_{H\otimes K} | \phi_H \rangle_{\mathfrak H} = 0$ whenever $\hat\rho_H |\phi_H\rangle = \hat\rho_H |\phi'_H\rangle = 0$. Denote now $P_{\hat\rho_H}$ the projector on the support of $\hat\rho_H$, $P_{\hat\rho_H}\hat\rho_H =\hat\rho_H P_{\hat\rho_H}=\hat\rho_H$, rewrite this property as

\[
\hat\rho_{H\otimes K} = P_{\hat\rho_H} \hat\rho_{H\otimes K} P_{\hat\rho_H}
\]   

\noindent and introduce the result in Eq.(\ref{eq15}) to obtain

\begin{equation}
\label{eq16}
Tr_H\left[{P_{\hat\rho_H}\Gamma^{(H)}\left({ Tr_K \hat\rho_{H\otimes K} }\right) P_{\hat\rho_H}\cdot \hat\rho_{H\otimes K} }\right] = 0\;.
\end{equation}    

\noindent Since $\hat\rho_{H\otimes K}$ is arbitrary, it follows that a necessary and sufficient condition for the complete positivity of the Polchinski extension of a dynamics of type (\ref{eq4}) is that the self-adjoint "dissipative part" $\Gamma^{(H)}$ of the generator has null action on the support of the local density matrix, that is

\begin{equation}
\label{eq17}
P_{\hat\rho_H}\Gamma^{(H)}\left({ Tr_K \hat\rho_{H\otimes K} }\right) P_{\hat\rho_H} = 0\;.
\end{equation} 

\noindent But because $\Gamma^{(H)}$ enters the equation of motion (\ref{eq5}) for $\hat\rho_H$ via the anticommutator $\{\Gamma^{(H)}, \hat\rho_H\}$, Eq.(\ref{eq17}) implies that the corresponding contribution in the extended equation of motion reduces to

\[
i\; \{\Gamma^{(H)}, \hat\rho_{H\otimes K} \} = i\; (I_H - P_{\hat\rho_H} ) \Gamma^{(H)} P_{\hat\rho_H} \hat\rho_{H\otimes K} + i\; \hat\rho_{H\otimes K} P_{\hat\rho_H} \Gamma^{(H)} (I_H - P_{\hat\rho_H} ) 
\] 

\[
=\left[{ i\; (I_H - P_{\hat\rho_H} ) \Gamma^{(H)} P_{\hat\rho_H} - i\; P_{\hat\rho_H} \Gamma^{(H)} (I_H - P_{\hat\rho_H} )\;,\; \hat\rho_{H\otimes K} }\right]
\]

\noindent In other words, the "dissipative" $\Gamma^{(H)}$ produces an action equivalent to that of an additional non-dissipative term, and is non-essential. This completes our proof.\\

Condition (\ref{eq17}) implies that nonlinear dynamics supporting completely positive Polchinski extensions cannot produce a variation in time of the eigenvalues of $\hat\rho_H$, and therefore enforce a nonlinear evolution of the eigenvectors with unitary propagators

\[
S_{\hat\rho_H}^{(H)} (t) \cdot \left[{S_{\hat\rho_H}^{(H)} (t)}\right]^{\dagger} = \left[{S_{\hat\rho_H}^{(H)} (t)}\right]^{\dagger} \cdot S_{\hat\rho_H}^{(H)} (t) = I_H
\]

\noindent The extended form of Eq.(\ref{eq5}), reading

\begin{equation}
\label{eq18}
i\hbar \dot{\hat\rho}_{H\otimes K} = \left[{ T^{(H)}\left({Tr_K\hat\rho_{H\otimes K}}\right), \hat\rho_{H\otimes K} }\right] + i \left\{ { \Gamma^{(H)}\left({Tr_K\hat\rho_{H\otimes K}}\right), \hat\rho_{H\otimes K} }\right\}\;\;,
\end{equation}

\noindent displays the same property, and preserves the eigenvalues of the total density matrix $\hat\rho_{H\otimes K}$. Hence both the total entropy and each of the local entropies remain stationary throughout an evolution that takes entangled states continuously into entangled states. This fact may be considered a good indication that the dynamics (\ref{eq18}) preserves the "degree of entanglement", or at least does not produce a pathological increase or discontinuity of entanglement, hence provides a physically acceptable entanglement dynamics. Moreover, from the particular case when the initial total density matrix is a rank 1 projector for a pure state, it can be seen that entangled pure states evolve nonlinearly into entangled pure states. 

A note is in order now regarding the following immediate corollary of {\bf Theorem 2}:\\

{\it {\bf Corollary (Theorem 2)}: Nonlinear equations of motion of type (\ref{eq5}) with essential dissipative contributions do not admit completely positive Polchinski extensions.}\\

A good example of an essentially dissipative nonlinear evolution is provided by the entropic dynamics of refs.\cite{EQD,Beretta-1,Beretta-2}. The local form of this dynamics yields a time-dependent spectrum of the density matrix [and entropy], and the associated $\Gamma^{(H)}$ does not comply with Eq.(\ref{eq17}), hence a completely positive Polchinski extension is not possible. Nevertheless, it is important to keep in mind that the absence of a Polchinski extension {\em does not imply} the absence of {\em any }completely positive extension for this kind of nonlinear dynamics. The Polchinski ansatz for the corresponding total propagator prescribes only a specific functional form, and by no means exhausts the continuum of possible completely positive extensions with proper entanglement dynamics. The correct meaning of the above {\bf Corollary }is that all well-behaved completely positive extensions for such cases are necessarily of a pseudo-nonseparable/nonlocal form [for a class of dynamics with well-behaved nonseparable extensions, albeit beyond the framework of Eq.(\ref{eq5}), see also Sec.6]. It must be pointed out that the apparent nonseparability/nonlocality referred to here is stronger then that discussed \cite{Czachor-1} in the framework of nonlinear von Neumann equations, where the total propagators are of the separable Polchinski type. 

For a simple example of a nonlinear dynamics with a non-essential $\Gamma^{(H)}$ that admits a Polchinski extension, satisfies the {\it pure state condition}, and produces a well-behaved entanglement dynamics, consider again the equation of motion (\ref{eq5}) with

\begin{equation}
\label{eq19}
T^{(H)}(\hat\rho_H) = H\hat\rho_H^q + \hat\rho_H^q H\;,
\end{equation}

\noindent and

\begin{equation}
\label{eq20}
\Gamma^{(H)}(\hat\rho_H) = (I_H - \hat\rho_H^{r-1}) A^{(H)} (I_H - P_{\hat\rho_H} ) + h.c. \;\;,
\end{equation}

\noindent where $r > 1$ is a real scalar, and $A^{(H)} \in L({\mathfrak H})$ is a self-adjoint operator. The explicit form of the extended dynamics (\ref{eq18}) in the presence of a passive, noninteracting environment reads

\[
i \hbar \;\dot{\hat\rho}_{H\otimes K} = \left[{  H ( Tr_K\hat\rho_{H\otimes K})^q +  (Tr_K\hat\rho_{H\otimes K})^q H\;, \;\hat\rho_{H\otimes K} }\right] + i\; \{ (I_H - \hat\rho_H^{r-1}) A^{(H)} (I_H - P_{\hat\rho_H} ) + h.c. \;, \;\hat\rho_{H\otimes K} \}\;
\]

\noindent and may be checked to be positive [e.g., by construction from the equation of motion for the {\it state operator }$\hat\gamma_{H\otimes K}$, $\hat\gamma_{H\otimes K}\hat\gamma_{H\otimes K}^\dagger = \hat\rho_{H\otimes K}$ ], trace preserving, separable, local for both the system $\mathfrak H$ and the environment $\mathfrak K$, and to evolve [pure] entangled states continuously into [pure] entangled states. Both the total entangled state and each of the reduced local states have eigenspectra stationary in time, and thus the total entropy and the local entropies are constants of motion. As already noted, one may infer from this that the "degree of entanglement" is also preserved.

\section{The measurement problem and nonlinear quantum dynamics}
\label{Sec5}

As stated in the introductory paragraph, self-consistent nonlinear theories must necessarily regard all mixed quantum states as {\it elementary mixtures}. A direct consequence of the "elementary mixture" interpretation is that a consistent description of quantum measurements in a nonlinear context must abandon both the use of the projection postulate, and the dynamical form of the "probabilistic mixture" interpretation \cite{qnlc-qd}. On the one hand, this conclusion simultaneously complements {\it and }supercedes popular objections \cite{nlin-ftl-1,nlin-ftl-2,nlin-ftl-3,Polchinski,Mielnik} that target, in one way or another, the dynamic "probabilistic" interpretation. On the other hand, one is confronted with a necessary reconsideration of the measurement problem in a nonlinear context. For nonlinear von Neumann theories with Polchinski-type extensions this issue was considered recently in ref.\cite{Czachor-Doebner}. There it is proposed that the Zeno effect in correlation measurements on entangled particles be described formally via "switching-off functions" that replace the "projections-at-a-distance" of the linear formalism. Then conditional probabilities in sequential measurements can be calculated, as usual, through products of quantum amplitudes along a selected "path" of possible sequential outcomes. This fundamental prescription remains valid particularly because Polchinski-type theories preserve a separable propagator form of the equations of motion for density matrices [see Eqs.(\ref{eq8}) and (\ref{eq13})]. 

We wish to add here a few explanatory comments regarding the problem of correlation probabilities and the physical basis for the concept of "switching-off functions". As in linear quantum theory, one may assume that a measurement operation leaves a quantum system in a convex superposition of projected "states", although a "probabilistic mixture" interpretation [not to be confused with the "static equivalence" interpretation] does not apply, and individual projected terms cannot bear independent reality. Also, the traces of the projected components retain their usual significance as probabilities of measurement outcomes. For example, a measurement that verifies the occurrence of pure states in the support of a projector $P_H = P_H^2 \in L_+ ({\mathfrak H})$ realizes the usual map

\begin{equation}
\label{eq21}
\hat\rho_H \to P_H \hat\rho_H P_H + Q_H \hat\rho_H Q_H \;\;,
\end{equation}    

\noindent where $Q_H =I_H - P_H \in L_+ ({\mathfrak H})$ is the complement of $P_H$, and the probability of a positive outcome for $P_H$ is given by $p(P_H) = Tr_H (P_H \hat\rho_H P_H)$. 

Suppose now that the post-measurement state evolves nonlinearly according to an equation of motion of type (\ref{eq5}). It is commonly believed that any nonlinear evolution will render the evolutions of individual projected components unseparable, and therefore will not be able to describe exact correlations under further measurements in a "natural way". This is a misperception. In reality, the projected components of the density matrix may or may not evolve independently of each other under an evolution of type (5), depending on whether the propagator $S^{(H)}$ leaves or not invariant the corresponding subspaces, in the sense that

\[
S^{(H)}_{P_H \hat\rho_H P_H + Q_H \hat\rho_H Q_H} = P S^{(H)}_{P_H \hat\rho_H P_H} P +  Q S^{(H)}_{Q_H \hat\rho_H Q_H} Q\;\;.
\]  

\noindent Apart from the nonlinear dependence of the generators on the density matrices, this behaviour is in perfect analogy with the standard theory, and also can be related directly to the fundamental symmetries of the dynamics. For a constructive proof that subspace invariance is indeed possible under a nonlinear dynamics, consider the example described by Eqs.(\ref{eq19}-\ref{eq20}) in the situation when the operators $H$ and $A^{(H)}$ leave the $P_H$ and $Q_H$ subspaces invariant, i.e., $H = P_H H P_H +Q_H H Q_H$ and $A^{(H)} = P_H A^{(H)} P_H + Q_H A^{(H)} Q_H$. It is easy to check that under such a prescription any density matrix of the form $P_H \hat\rho_H P_H + Q_H \hat\rho_H Q_H$ develops independent evolutions for the projected components $P_H \hat\rho_H P_H$ and $Q_H \hat\rho_H Q_H$, so that 

\begin{equation}
\label{eq22}
S^{(H)} (P_H \hat\rho_H P_H + Q_H \hat\rho_H Q_H) \left[{ S^{(H)} }\right]^\dagger = S^{(H)}_P (P_H \hat\rho_H P_H) \left[{ S^{(H)}_P }\right]^\dagger + S^{(H)}_Q (Q_H \hat\rho_H Q_H) \left[{ S^{(H)}_Q }\right]^\dagger \;\;.
\end{equation}

\noindent Here the propagator notation has been simplified as $S^{(H)} = S^{(H)}_{P_H \hat\rho_H P_H + Q_H \hat\rho_H Q_H}$, $S^{(H)}_P = P S^{(H)}_{P_H \hat\rho_H P_H} P$ and $S^{(H)}_Q = Q S^{(H)}_{Q_H \hat\rho_H Q_H} Q$. We note also that if $P_H$ singles out a pure state, then the post-measurement propagation evolves this pure state in a {\it linear, Hamiltonian }way. 

Again as in the linear theory, the two-measurement correlation problem for this kind of nonlinear dynamics, with adequate symmetries, does have a "natural" solution. Let us assume that the post-measurement projected components evolve indeed independently. Suppose at a latter time a measuring device is coupled exclusively to one of the components, say $P_H$, and performs a second measurement of the type $P_H$ vs. $Q_H$. According to the measurement map (\ref{eq20}), both the state of the system and the probability of a positive result for $P_H$ will remain unchanged. It follows that the correlation probability between a positive outcome for the first measurement and a positive outcome for the second measurement remains unit, as one is used to obtain in linear theory. If $P_H$ represents a pure state, we recover the usual correlation result for pure states. 

Moreover, a very interesting situation arises when all post-measurement projected components correspond to pure states and each of these states happens to be a stationary state of the post-measurement nonlinear evolution. In the specific example of Eqs.(\ref{eq19}-\ref{eq20}), this corresponds to the case when $[H, A^{(H)}] = 0$ and the measurement projects on a complete set of common eigenstates of $H$ and $A^{(H)}$. Then the entire post-measurement density matrix is stationary and any evidence of a nonlinear evolution vanishes. All observations from the second measurement are automatically confined to the linear limit of the theory, i.e. to standard quantum physics. This conclusion becomes particularly intriguing in the case of two-state [spin-1/2] systems, when all orthogonal measurements involve only two pure states. 

Consider now the problem of correlation measurements on entangled systems. Given that "probabilistic mixtures" and the "projection postulate" are incompatible with a nonlinear dynamics, local measurements on entangled systems can only be assigned the same status as other local interactions. While this point of view is optional under a linear dynamics, here it becomes an indispensable premise. It implies that local measurements must be described exclusively in terms of completely positive local operations in the sense of {\bf Definition 2}, and necessarily leave the reduced states of remote entangled counterparts unaffected. It also implies that incomplete sets of local projections do not qualify as proper measurement descriptors. Only complete sets of projections, which generate completely positive projective operations [i.e., trace-preserving, completely positive operations in the linear sense], can be associated with physical measurements and events. As in other "no-projection" theories, the emerging overall philosophy is that probabilities can be calculated {\it as if }measurements project the total state, but the projections themselves retain just a virtual significance, and cannot be considered genuine physical events. According to this reasoning, correlation probabilities for two-measurement experiments are given by the usual trace rule over the proper {\it virtual history} of the system through the measurement setup.  
 
A direct corollary following from the complete positivity of local measurements and the "trace over virtual histories" rule is that correlation probabilities for an entangled dynamics described by a Polchinski extension can be calculated with the "switch-off" rule of ref.\cite{Czachor-Doebner}, although the exact trace rule takes into account the complete evolutions. Indeed , assume a Polchinski extension such that the total equation of motion, of nonlinear von Neumann type, reads

\begin{equation}
\label{eq23}
i \hbar \;\dot{\hat\rho}_{H\otimes K} = \left[{ T^{(H)}_{Tr_K \hat\rho_{H\otimes K}} +  T^{(K)}_{Tr_H \hat\rho_{H\otimes K}}\;,\; \hat\rho_{H\otimes K} }\right]\;\;.
\end{equation}

\noindent A local $\mathfrak H$ measurement of the $P_H$ vs. $Q_H$ type, performed at time $t_1$, generates the total state

\[
\hat\rho_{H\otimes K}(t_1+0) = P_H\hat\rho_{H\otimes K}(t_1) P_H + Q_H \hat\rho_{H\otimes K}(t_1)Q_H \;\;.
\]

\noindent Assume also that, after the measurement, the $\mathfrak H$ dynamics leaves the $P_H$ and $Q_H$ subspaces invariant such that

\begin{equation}
\label{eq24}
T^{(H)}_{Tr_K \hat\rho_{H\otimes K}} \to T^{(H)}_P + T^{(H)}_Q\;,
\end{equation}

\noindent where for economy we have denoted $T^{(H)}_P = P_H T^{(H)}_{P_H (Tr_K \hat\rho_{H\otimes K}) P_H} P_H$ and $T^{(H)}_Q = Q_H T^{(H)}_{Q_H (Tr_K \hat\rho_{H\otimes K}) Q_H} Q_H$ [see also the explicit example given earlier in this Sec.]. On the other hand, the local $\mathfrak K$ generator transforms as 

\begin{equation}
\label{eq25}
T^{(K)}_{Tr_H \hat\rho_{H\otimes K}} \to T^{(K)}_{Tr_H [ P_H \hat\rho_{H\otimes K} P_H + Q_H \hat\rho_{H\otimes K} Q_H ] } \equiv T^{(K)}_{Tr_H \hat\rho_{H\otimes K}}\;\;,
\end{equation}

\noindent i.e., is not affected by the change of local state and dynamics in the $\mathfrak H$ system. This confirms that the local $\mathfrak K$ dynamics does not experience any discontinuities that may allow a local observation of the remote $\mathfrak H$ measurement. In agreement with refs.\cite{Czachor-1,Czachor-2,Czachor-6,Czachor-Doebner}, we must stress that this conclusion is precisely contrary to the common assumptions advocated in popular arguments against nonlinear quantum dynamics \cite{nlin-ftl-1,nlin-ftl-2,nlin-ftl-3,Polchinski}. Under these conditions, the total post-measurement dynamics, for $t > t_1$, takes the form 

\[
\hat\rho_{H\otimes K} (t \equiv  P_H \hat\rho_{H\otimes K} (t) P_H + Q_H \hat\rho_{H\otimes K}(t) Q_H  
\]

\begin{equation}
\label{eq26}
S^{(K)}_{Tr_H \hat\rho_{H\otimes K}}(t) S^{(H)}_P (t)  P_H \hat\rho_{H\otimes K}(t_1) P_H  \left[{S^{(H)}_P (t)  S^{(K)}_{Tr_H \hat\rho_{H\otimes K}}(t) }\right]^{\dagger} + S^{(K)}_{Tr_H \hat\rho_{H\otimes K}}(t) S^{(H)}_Q(t)  Q_H \hat\rho_{H\otimes K}(t_1)Q_H  \left[{S^{(H)}_Q(t)  S^{(K)}_{Tr_H \hat\rho_{H\otimes K}}(t) }\right]^{\dagger} \;,
\end{equation}

\noindent where $S^{(H)}_P$, $S^{(H)}_Q$, and $S^{(K)}_{Tr_H \hat\rho_{H\otimes K}}$ denote the unitary propagators corresponding to $T^{(H)}_P$, $T^{(H)}_Q$, and $T^{(H)}_{Tr_K \hat\rho_{H\otimes K}}$, respectively. Now let a second measurement be performed locally on the $\mathfrak K$ system at a time $t_2 > t_1$, say to probe a subspace projected by $P_K = P_K^2$ vs. the complement projected by $Q_K = I_K - P_K$. The new post-measurement state becomes

\[
\hat\rho_{H\otimes K}(t_2+0) = P_K \hat\rho_{H\otimes K}(t_2) P_K + Q_K \hat\rho_{H\otimes K}(t_2)Q_K
\] 

\begin{equation}
\label{eq27}
= P_K S^{(K)}_{Tr_H \hat\rho_{H\otimes K}}(t_2) S^{(H)}_P(t_2)  P_H \hat\rho_{H\otimes K}(t_1) P_H  \left[{S^{(H)}_P (t_2) }\right]^{\dagger} \left[{ S^{(K)}_{Tr_H \hat\rho_{H\otimes K}}(t_2) }\right]^{\dagger} P_K + \;etc.
\end{equation}

\noindent As under a linear dynamics, the trace of the term shown explicitly on the second line above provides the joint probability that a positive $P_H$ measurement is followed by a positive $P_K$ measurement. Due to the unitarity of the propagators, this unnormalized probability can be cast as

\[
p(P_K|P_H) = Tr\left[{ P_K S^{(K)}_{Tr_H \hat\rho_{H\otimes K}}(t_2) S^{(H)}_P(t_2)  P_H \hat\rho_{H\otimes K}(t_1) P_H  \left[{S^{(H)}_P (t_2) }\right]^{\dagger} \left[{ S^{(K)}_{Tr_H \hat\rho_{H\otimes K}}(t_2) }\right]^{\dagger} P_K }\right] 
\]

\begin{equation}
\label{eq28}
= Tr \left[{P_K S^{(K)}_{Tr_H \hat\rho_{H\otimes K}}(t_2) P_H \hat\rho_{H\otimes K}(t_1) P_H \left[{ S^{(K)}_{Tr_H \hat\rho_{H\otimes K}}(t_2) }\right]^{\dagger} P_K }\right].
\end{equation}

\noindent However, a more symmetric form is obtained if the total state at time $t_1$ is expressed as the evolved of a state at a previous time $t_0$, e.g., 

\begin{equation}
\label{eq29}
\hat\rho_{H\otimes K}(t_1) = S^{(K)}_{Tr_H \hat\rho_{H\otimes K}}(t_1) S^{(H)}_{Tr_K \hat\rho_{H\otimes K}}(t_1) \hat\rho_{H\otimes K}(t_0)  \left[{S^{(H)}_{Tr_K \hat\rho_{H\otimes K}}(t_1)}\right]^{\dagger} \left[{S^{(K)}_{Tr_H \hat\rho_{H\otimes K}}(t_1)}\right]^{\dagger}.
\end{equation}

\noindent When this expression is substituted into the joint probability (\ref{eq28}), the product of $\mathfrak K$ propagators between $t_0$ and $t_1$, and $t_1$ and $t_2$, respectively, produces the unperturbed propagator between $t_0$ and $t_2$, since the evolution of the $\mathfrak K$ system remains continuous in this time interval. In this way one finds that

\begin{equation}
\label{eq30}
 p(P_K|P_H)= Tr \left[{P_K P_H S^{(K)}_{Tr_H \hat\rho_{H\otimes K}}(t_2) S^{(H)}_{Tr_K \hat\rho_{H\otimes K}}(t_1) \hat\rho_{H\otimes K}(t_0)\left[{S^{(H)}_{Tr_K \hat\rho_{H\otimes K}}(t_1)}\right]^{\dagger} \left[{ S^{(K)}_{Tr_H \hat\rho_{H\otimes K}}(t_2) }\right]^{\dagger} P_H P_K }\right] \;\;.
\end{equation}

\noindent But let us observe that the effective compound propagator $S^{(K)}_{Tr_H \hat\rho_{H\otimes K}}(t_2) S^{(H)}_{Tr_K \hat\rho_{H\otimes K}}(t_1)$ in the expression above corresponds to an effective evolution driven by a total generator of the piecewise form

\begin{equation}
\label{eq31}
T_{eff}^{(H\otimes K)} (t) = \theta (t_1-t) \; T^{(H)}_{Tr_K \hat\rho_{H\otimes K} } + \theta (t_2-t) \; T^{(K)}_{Tr_H \hat\rho_{H\otimes K} }\;,
\end{equation}

\noindent where $\theta(t) =1$ for $t < 0$ and $\theta(t) =0$ otherwise. With this convention, Eq.(\ref{eq31}) reproduces precisely the prescription of "switching-off functions" employed in ref.\cite{Czachor-Doebner}. We conclude by emphasizing that in contrast to ref.\cite{Czachor-Doebner}, the above derivation did not require a modification of the Polchinski extension for entangled dynamics. Aside from the standard trace rule for joint probabilities, it relies on the assumptions that measurements are completely positive operations, and that the post-measurement dynamics leaves the tested subspaces invariant.

\section{Additional classes of well-behaved nonlinear dynamics}
\label{Sec6}
  
The class of nonlinear extensions outlined in Secs.4.2-4.3 is far from exhaustive. For instance, a straightforward generalization can be obtained from a convex linear superposition [of a finite number] of such processes, by defining
\begin{equation}
\label{eq200}
\hat\rho(t) = \sum \limits_k {\lambda_k S_{\hat\rho_k}^{(k)}(t) \hat\rho(0) \left[{ S_{\hat\rho_k}^{(k)} (t)}\right]^\dagger}\;,
\end{equation}
\noindent for $\lambda_k > 0$. Here the lower label $\hat\rho_k$ means that the propagator $S^{(k)}$ is to be understood as generated exclusively by process $k$ if starting from the initial state $\hat\rho(0)$ [in effect $\hat\rho(t) = \sum\limits_k {\lambda_k \hat\rho_k (t)}$]. 

Evidently, if every individual process admits a well-behaved completely positive extension, conserves probability [and energy], and propagates pure states in a linear manner, the total process also will display the same properties and will qualify as a physically meaningful dynamics. Note that if the individual completely positive extensions are of the explicitly separable Polchinski type, the overall extension is no longer explicitly separable. This is in agreement with the fact that the total dynamics is essentially dissipative, although it falls outside the scope of {\bf Theorem 2}. 

Also note that pure states need not propagate unitarily in this case, since individual processes may contribute distinct linear generators in the pure state limit. For example, one may take nonlinear unitary processes generated by $T_k(\hat\rho_k)=H_k\hat\rho_k^{q_K} + \hat\rho_k^{q_k} H_k$ and $\Gamma_k = 0$, where the scalars $q_k > 0$ and the linear, hermitian operators $H_k$ are generally distinct for every $k$. Then each process propagates pure states unitarily into pure states with a different "Hamiltonian" $H_k$, but the total pure state propagator is no longer unitary, unless $H_k = H, \forall k$. 

\section{Summary}
\label{Sec7}

We have presented a novel framework for the construction of trace-preserving and positive definite nonlinear quantum dynamics, based on the square-root decomposition of a density matrix. The class of nonlinear dynamics obtained in this approach generalizes the previously known class of nonlinear von Neumann equations \cite{Czachor-6}, which are retrieved as a particular case. In addition, we have provided a compact definition of completely positive, relativistically well-behaved nonlinear extensions to entangled systems. In contrast to the linear case, such extensions are not unique, and different completely positive extensions generate different entanglement dynamics. For the selection of a unique extension it is necessary to refer to an additional physical [entanglement] or functional criterion . In this sense, we showed that explicitly separable Polchinski extensions are possible if and only if the local dynamics is not essentially dissipative, i.e. can be brought to a nonlinear von Neumann form. Conversely, essentially dissipative dynamics must be expected to involve completely positive propagators that, although local and separable, do not have a manifestly separable functional form on entangled states. 

In view of our argument in ref.\cite{qnlc-qd} , we paid particular attention to the possibility that some nonlinear dynamics may evolve {\it all pure states }of isolated systems unitarily into pure states, as in linear quantum theory. We emphasized that dynamics that comply with this "pure state condition", but evolve mixed states in a nonlinear manner, do indeed exist, and support relativistically well-behaved [i.e., nonlinear completely positive] extensions to entangled systems. 

We have also considered the issue of correlation measurements under a nonlinear dynamics, and brought additional support to the idea \cite{Czachor-Doebner} that nonlinear dynamics can be consistent with quantum measurement theory. Particularly, we showed that there are nonlinear dynamics that can reproduce in a natural way the familiar unit correlation probabilities in two-measurement set-ups. These dynamics also produce naturally the "switching-off " rules proposed in ref.\cite{Czachor-Doebner} for two-measurement experiments on entangled systems. Our fundamental assumptions are that quantum measurements must be described by completely positive operations, and that correlation probabilities are calculated by the usual trace rule applied to virtual measurement histories.

\end{document}